# Purcell-enhanced X-ray scintillation

Yaniv Kurman[1,2,†], Neta Lahav[1,2,3,†], Roman Schuetz[1,2,†], Avner Shultzman[1,2], Charles Roques-Carmes[4], Alon Lifshits[1], Segev Zaken[1], Tom Lenkiewicz[1,2], Rotem Strassberg[2], Orr Be'er[3], Yehonadav Bekenstein[2,3] and Ido Kaminer[1,2]

[1]Department of Electrical and Computer Engineering, Technion - Israel Institute of Technology, 32000 Haifa, Israel
[2]Solid state institute, Technion - Israel Institute of Technology, 32000 Haifa, Israel
[3]Department of Materials Science and Engineering, Technion - Israel Institute of Technology, 32000 Haifa, Israel
[4]Research Laboratory of Electronics, Massachusetts Institute of Technology, Cambridge, MA 02139, USA
[†]These authors contributed equally to the manuscript

**Scintillation materials convert high-energy radiation to optical light through a complex multi-stage process. The last stage of the process is light emission via *spontaneous emission*, which usually governs and limits the scintillator emission rate and light yield. For decades, the quest for faster emission rate and greater light yield motivated the frontier of scintillators research to focus on developing better materials and dopants. Here, we experimentally demonstrate a fundamentally different, recently proposed concept for enhancing the scintillation rate and yield: the *Purcell effect*. The Purcell effect is a universal enhancement mechanism for spontaneous emission by engineering the optical environment. In scintillators, such an enhancement arises from engineering the nanoscale geometry within the scintillation bulk, which thus applies universally to any scintillating material and dopant. We design and fabricate a thin multilayer nanophotonic scintillator, demonstrating Purcell-enhanced scintillation, achieving a 50% enhancement in emission-rate and an 80% enhancement in light yield. We demonstrate the potential of our device for realizing these enhancements in real-life settings for X-ray applications, also due to the robustness of the nanophotonic design to fabrication disorder. Our results show the bright prospects of bridging nanophotonics and scintillators science, toward reduced radiation dosage and increased resolution for high-energy particles detection.**

**Introduction**

Scintillators are at the heart of a wide range of technologies, from X-ray-based computerized tomography and positron-emission-tomography[1,2] to advanced detectors in particle physics[3]. Although these applications span a wide range of energy and length scales, they are all based on the same underlying mechanism of scintillation, in which high-energy radiation triggers the emission of optical photons that are more easily detected[4]. Throughout the years, major developments in scintillator science led to the creation of efficiently emitting transparent materials with sufficient stopping power for various high-energy particles[5,6]. A long-lasting and still ongoing challenge is to find materials that convert the energy provided by the stopped particles to visible photons with high emission rate and yield[7,8]. This conversion is greatly dependent on the scintillating material properties and on the particle type and energy. Nevertheless, the final stage of the conversion process is common to all scintillators, regardless of the material constituents and the detected particle: the output optical photons are created via *spontaneous emission*.

The nature of spontaneous emission imposes critical limitations on all scintillators. First, the spontaneous emission rate, which is assumed intrinsic to each scintillation material, sets an upper limit on the image frame rate[1], and even the spatial resolution of medical imaging applications such as positron emission tomography[9,10]. Second, the emission in bulk scintillators is intrinsically isotropic, which often causes as high as 70% of the photons to be emitted into undetectable angles (e.g., below total internal reflection)[11], limiting scintillators' light yield (Fig. 1a, top panel).

Different solutions were proposed for bypassing each of the limits that hinder the operation of scintillators. For example, emission into undetectable angles can be captured by structuring the scintillator into pixels with transverse (metallic) boundaries that reflect photons to detectable directions[11,12]. However, this solution prolongs the detection time and is inapplicable in many practical cases. Instead, a more recent approach involves structuring the surfaces of scintillators into photonic crystals[13–15], improving the out-coupling of the emitted photons and thus increasing the overall light yield. This approach, which combines nanophotonic structures on top of scintillators, has been coined *nanophotonic scintillators*. Nevertheless, none of those approaches has managed to modify the intrinsic spontaneous emission nature of the bulk material, which lies at the heart of all scintillation processes.

Taking a different approach to nanophotonic scintillators, we recently proposed the concept of *Purcell-enhanced scintillators*[16], wherein the predetermined nanoscale geometry of the scintillator volume controls its intrinsic spontaneous emission properties. This novel type of nanophotonic scintillators consists of nanoscale structuring of the scintillation *bulk*, combining materials of alternating index of refraction (in one, two, or three dimensions). We predicted that a precise nanostructure design could increase the emission rate and shape the angular distribution of spontaneous emission in scintillators, resulting in greater directionality and faster emission, by leveraging an angle-dependent Purcell effect. Recent work further enriched the ideas for such unique types of nanophotonic scintillators[17,18].

Since the X-ray beam is illuminating a large volume, the design strategy to enhance the local density of photonic states (LDOS) over the scintillation volume is different from a conventional single-emitter Purcell enhancement scenario. Therefore, the enhancement depends on spatial coordinates over dimensions greater than the wavelength, and its angular distribution can be made non-uniform. Our approach is completely material agnostic and can be applied to both well-established[5] and new scintillator material platforms[8] as long as their nanoscale geometry can be controlled.

In this paper, we present the first experimental realization of Purcell-enhanced scintillation. This work also provides the first observation of the Purcell effect in a new regime: namely, when dipolar transitions in a nanophotonic structure are induced by an X-ray excitation, i.e., the X-ray-driven Purcell effect. We demonstrate enhanced emission rate and yield following an X-ray excitation of a thin nanophotonic scintillator. These enhanced capabilities arise from a specially designed nanophotonic structure prepared by alternating layers of a conventional scintillator (europium- and bismuth-doped lutecia doped, $Lu_2O_3$:Eu-Bi) and a conventional dielectric (silica, $SiO_2$). This designed multilayer structure imparts an angle-dependent Purcell enhancement, which we measure and compare for both UV excitation (photoluminescence) and X-ray excitation (scintillation). The effect is reflected in enhanced spontaneous emission rates into desired directions and inhibited emission into undesired directions. Instead of being emitted isotropically (Fig. 1a, top panel), the generated photons are emitted to useful angles upon creation (Fig. 1a, bottom panel).

The Purcell effect in our nanophotonic scintillator enhances the scintillation rate of emission (Fig. 1b) – a result that had never been achieved for X-ray-induced processes. This result is especially intriguing because the scintillation emission rate is often regarded as an intrinsic property of the material and is now instead shown to be modified by the geometry. Most intriguingly, our design shows tolerance to fabrication imperfections which we show originates from an intrinsic robustness of the photonic design. We stress that the same concept could be implemented with any scintillator material and could be generalized to additional geometries.

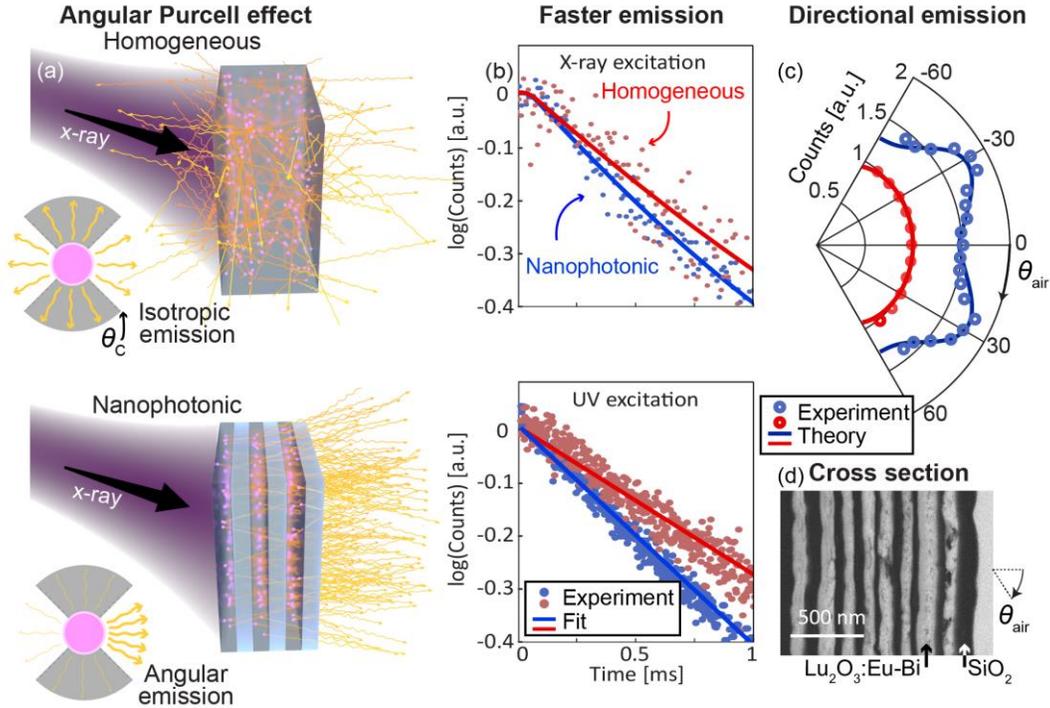

**Figure 1: Purcell-enhanced scintillator: overview. (a)** In a conventional homogeneous scintillator (top), the energy of incoming X-rays (black arrow) is converted into isotropically emitted visible photons (yellow arrows), of which only a fraction reaches an external detector ($\theta_c$ denotes the critical angle). A nanophotonic structure composed of scintillating materials (bottom) can emit photons mostly into detectable angles and thus increase the scintillation signal. **(b)** Measurement of emission timing under X-ray (top) and UV (bottom) excitation, both measured at an angle of 30°. In both plots, emission from the homogeneous thin-film scintillator (resp. nanophotonic scintillator) is shown in red (resp. blue). The linear fitting for both excitation types (solid curves) shows decay times of $1.635 \pm 0.008$ ms vs $1.107 \pm 0.004$ ms for the homogeneous scintillator vs the nanophotonic scintillator, respectively. **(c)** Angular emission in air ($\theta_{\text{air}}$) following X-ray excitation of a homogeneous (red) and nanophotonic (blue) scintillator, having the same volume of scintillating material. The angular-Purcell enhancement provides an overall increase of the collected light, in good agreement with the theoretical prediction (solid curves). **(d)** Cross-section image of the fabricated nanophotonic scintillator.

**Nanophotonic Purcell-enhanced scintillator characterization**

The underlying mechanism guiding the design of our nanophotonic scintillator is the control over the process of spontaneous emission using the Purcell effect. The conventional use of the Purcell effect is to enhance the spontaneous emission rate of a dipolar transition of an emitter placed inside an optical cavity[19] inside which the local density of photonic states (LDOS) is increased. A wide range of technologies are based on the Purcell effect, from lasers[20] and light emitting diodes[21] to single photon emitters[22]. In all of these, the optical cavity is designed to minimize the optical mode volume, V, and to maximize its quality-factor, Q, since the Purcell factor is proportional to $Q/V$[19].

Our recent theoretical proposal of nanophotonic Purcell-enhanced scintillation[16] shows that a completely opposite set of rules are required to enhance scintillation using the Purcell effect: a spatially extended optical mode with a low quality-factor. That is, the optimal design must allow emission from the entire scintillation volume into modes that efficiently couple out. This fundamental discrepancy originates in the physical nature of scintillation: many incoherent emitters distributed over the scintillation volume. Additional enhancement arises from specifically increasing the emission into desired angles and decreasing the emission into undesired angles. For that purpose, we design the optical structure according to its *angular-Purcell enhancement* instead of the conventional 'global' Purcell enhancement[16,23].

Previous measurements of Purcell enhancements were limited so far to electronically-driven[20,21], light-driven[22], and free-electron-driven experiments[24-26]. The latter two belong to the family of photoluminescence and cathodoluminescence effects. Examples of Purcell-enhanced cathodoluminescence are found when imaging plasmonic modes with high resolution[25] or probing the local photonic density of states[26]. All these experiments rely on electron excitations at the surface or in a shallow portion of the material near it, experiencing Purcell enhancements around localized points near the surface. In contrast, Purcell-enhanced scintillators are designed with nanoscale features over a larger volume, corresponding to the penetration depth required to stop energetic X-rays.

To demonstrate the above ideas and guidelines[16], we designed a thin nanophotonic structure from a scintillating material. Fig. 1d shows a cross section of the fabricated 16-layered structure. The nanophotonic structure is composed of alternating high-index scintillating material ($Lu_2O_3$:Eu-Bi) and low index spacers ($SiO_2$). When comparing the light yield of the thin nanophotonic scintillator to a homogeneous film scintillator ($Lu_2O_3$:Eu-Bi only) of the same scintillation material and same thickness, our nanophotonic structure shows up to 1.8-fold enhancement (Fig. 1c) for a large range of angles (see experimental setup in SM section S.2). Such 1.8 enhancement in the light yield translates to a $\sqrt{1.8}$ improvement in signal-to-noise ratio for any imaging and detection application.

Naturally, the thin thickness of the nanophotonic scintillator (1.3 μm) only stops a portion of the X-rays (8% in our case), and thus its absolute performance in terms of total quantum efficiency is inferior to many thick scintillators (discussed further in Fig. 4). However, we can estimate the performance of a thicker scintillator of the same design: Normalizing the number of scintillated photons by the total X-ray energy that was stopped by the structure shows that our thin Purcell-enhanced scintillator has an overall light yield of 26,000 photons/MeV, which surpasses a similarly normalized industrially used LYSO scintillator (details in SM section S.8).

We verify that this enhancement originates from the Purcell effect by measuring the scintillation decay curve using both X-ray and UV excitations (Fig. 1b). We show a reduction from decay times of $\tau_d = 1.635 \pm 0.008$ ms for the homogeneous scintillator to $\tau_d = 1.107 \pm 0.004$ ms for the nanophotonic structure at emission angle of 30° (SM sections S.5 and S.6 presents the UV and X-ray lifetime extraction, time-angle dependence, and full raw temporal measurements data in Figs. S11 and S12). Such reduction of lifetime is inherent to the Purcell effect and could be applied for faster scintillators such as LSO if fabricated in a nanophotonic structure. Shortening emission decay times is especially desirable for nuclear medicine applications since the imaging resolution, aka coincidence time resolution (CTR), follows CTR $\propto \sqrt{\frac{\tau_d}{\text{light yield}}}$, which our current design can directly improve by a factor of 1.6.

Beyond enhancement of the total light yield and decay time, scintillators with controllable angular properties can be desirable for high-resolution imaging applications. The nanophotonic scintillator structure not only enhances the total light yield, but also shapes its angular emission distribution at different frequencies, as evident in the calculated (Fig. 2a) and measured (Fig. 2b) angular-spectral distribution. This angular-dependence contrasts with the isotropic emission of the homogeneous scintillator (Fig. 2c). We correlate the overall stronger emission to faster emission rates by measuring the decay curves at two different emission angles (Fig. 2d), consistent with our prediction of an angle-dependent Purcell effect. Specifically, the emission rate enhancement is $1.5 \pm 0.02$ ms and $1.25 \pm 0.02$ ms at angles of 30° and 0°, respectively.

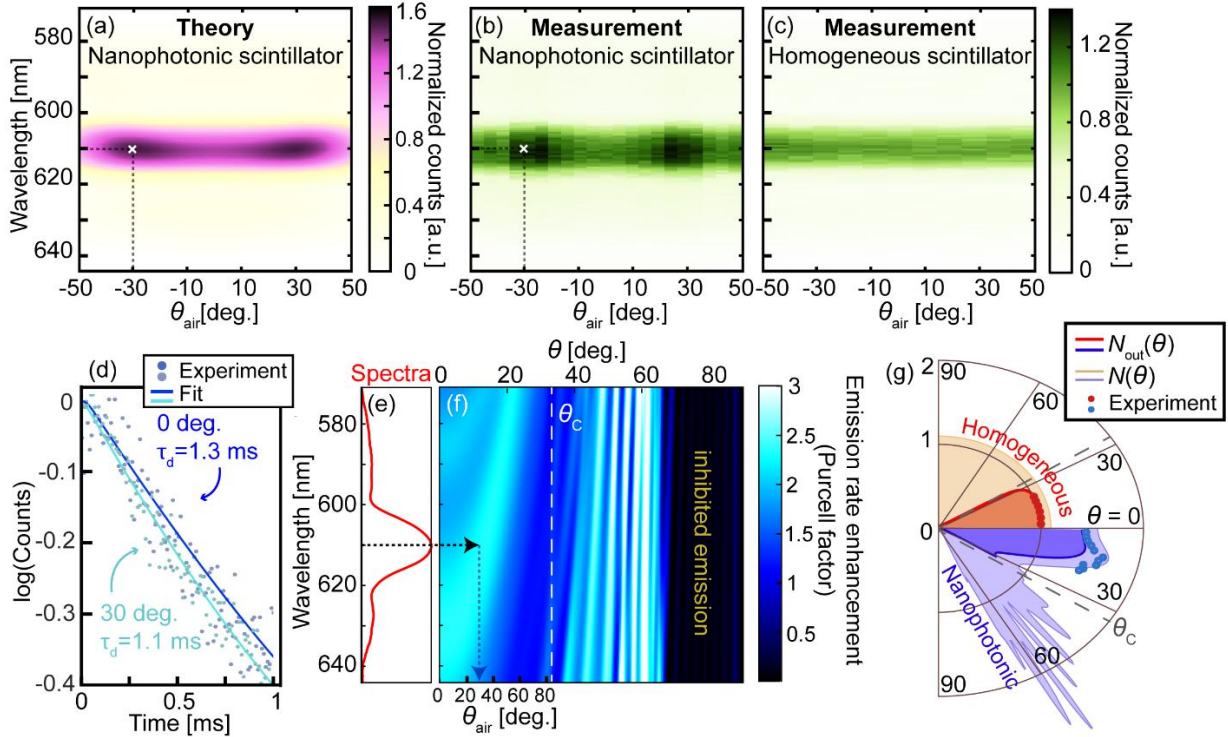

**Figure 2: Angular-Purcell enhancement in scintillation: theory and experiments**. **(a-c)** Light yield as a function of wavelength and emission angle in air $\theta_{\mathrm{air}}$ for an ideal nanophotonic scintillator (theoretical, a), the fabricated nanophotonic scintillator (experimental, b), and a homogeneous scintillator (experimental, c). The differences between (a) and (b) arise from structural imperfections (discussed in Fig. 3). **(d)** The nanophotonic scintillator decay curves at angles of 0° (blue) and 30° (cyan) under X-ray excitation, showing angular Purcell enhancement of 1.25±0.02 ms and 1.5±0.02 ms, respectively (calculated from the fit for their decay times, solid curves). **(e)** The measured scintillation emission spectra of $Lu_2O_3$:Eu. **(f)** Theory of emission rate enhancement (Purcell factor) for the nanophotonic scintillator as a function of wavelength and emission angle in air ($\theta_{\mathrm{air}}$, bottom $x$-axis) or inside the material ($\theta$, top $x$-axis). No emission occurs into modes that cannot propagate inside the photonic crystal (beyond ~70° inside the scintillator), thus increasing the number of photons that are emitted into detectable directions. Within the range of detectable angles, i.e., below the critical angle $\theta_C$ (left of the white dashed line), the emission is maximized at 30° in air (marked by arrows), which is consistent with our measurement. **(g)** The angular density of the scintillation emission $N(\theta)$ and outcoupled emission $N_{\mathrm{out}}(\theta)$ for the homogeneous film (upper half) and nanophotonic (lower half) scintillators, overlaid by dots to denote the experimental data. The nanophotonic design inhibits emission above 65°, enhancing in return the emission into detectable angles. Interestingly, the theory shows sharp resonances of enhanced emission into guided modes at 35°-65° that partially couple out via scattering (discussed in Fig. 3).

Figs. 2f,g elucidate the underlying physics by which our nanophotonic structure enhances the scintillation. Fig. 2f presents the theoretical Purcell factor of the photonic crystal, as a function

of wavelength and angle. The observed angular-dependence of the scintillation is determined by the photonic crystal band structure, which enhances the emission into a certain angle ($\theta_{\text{air}} = 30°$) at the peak wavelength of the $Lu_2O_3$:Eu emission spectra (612 nm, Fig. 2e), as shown by the dashed lines and the cross.

The Purcell effect also alters the scintillation emission at higher angles, beyond the critical angle. The emission manipulation includes (i) emission inhibition into angles that cannot propagate in the photonic crystal (above 65°), and (ii) enhanced emission into in-plane guided modes that may partially out-couple due to scattering by imperfections (35°-65°). We quantify these effects and the overall enhancement of the emission into the desired angles by calculating the number of emitted photons into each angle, as marked by the shaded area of Fig. 2g:

$$N(\theta) = N_{\text{tot}} \frac{\int d\lambda \, F_P(\theta, \lambda) Y(\lambda)}{\int d\theta \int d\lambda \, F_P(\theta, \lambda) Y(\lambda)}. \tag{1}$$

In the above equation, $\lambda$ is the emission wavelength, $Y(\lambda)$ is the emission spectra (Fig. 2e), $\theta$ is the emission angle inside the scintillator, and $N_{\text{tot}}$ is the total number of photons that are created within the scintillator. In addition, $F_P$ is the nanophotonic scintillator's Purcell factor[16,23]. The resulting angular spread increases the probability of emission towards the photodetector because fewer undetectable photons are created per X-ray excitation, leaving more photons to be emitted toward the detectable angles. We note that $N_{\text{tot}}$ is a conserved quantity that only depends on the scintillation material and the stopped X-ray energy (as long as the enhanced spontaneous emission into free-space photons is not becoming dominated by other non-radiative processes).

We can quantify the measured angular spread that couples out using $N_{\text{out}}(\theta) = N(\theta) T(\theta)$, where $T(\theta)$ is the transmission coefficient of the nanophotonic structure[16]. Both the transmission and the emission rate enhancement are angular dependent and are incorporated in our formalism. This calculation matches well with the measured data, showing that the nanophotonic scintillator light yield overcomes that of the homogeneous film (thick lines in Fig. 2g) as a result of an angular Purcell enhancement that controls the emission direction. Such a geometry-related enhancement mechanism fundamentally differs from the conventional mechanisms studied in the field of scintillators that are all material-related, altering $N_{\text{tot}}$ rather than the integral factors in Eq. (1).

Since the light is collected from a large area, we consider a Purcell factor averaged over the emitters' spectrum and locations, creating an effective Purcell factor. We notice that the Purcell factor in a multi-layer structure $F_p(z, q, \lambda)$ is defined for a specific location in space $z$, in-plane momentum $q$ (which is connected to the angle of emission), and wavelength $\lambda$, as calculated in Ref. [16]. Then, in order to examine a whole structure, we define the effective Purcell factor for each frequency and in-plane momenta as:

$$F_{p,\text{eff}}(q) = \int d\lambda Y(\lambda) \int dz G(z) \int_{\text{scint}} dz\, F_p(z, q, \lambda) \tag{2}$$

where the z integration goes over all scintillation locations with a density profile of $G(z)$, and $Y(\lambda)$ describes the emission spectra, shown in Fig. 2f. Importantly, the local Purcell factor and thus also the effective Purcell factor in Eq. (2) are strongly dependent on the thicknesses and refractive indices of each layer of the structure.

The angular dependence of the Purcell effect underpins our experimentally observed phenomena, showing angular dependence of both the emission *intensity* and *rate*. Interestingly, these phenomena are fundamentally different: while angular dependence of the emission *intensity* could arise from just patterning the surface (e.g., an angular filter), the angular dependence of the emission *rate* can only arise from the Purcell effect in a *volume* of emitters. We expect this latter effect to be especially ubiquitous in the X-ray driven Purcell effect, because X-rays entail the excitation of a large, distributed volume of emitters. In contrast, under optical excitations, the Purcell effect's angular dependence is less ubiquitous, although still documented in nanophotonics[27-30].

At the core of this novel phenomenon of angular-dependent emission rate is the fact that any scintillator is formed as an ensemble of microscopic emitters, where each can have a specific angular distribution for its emission, due to its nanophotonic environment. Once averaging over the ensemble of emitters, the overall emission will showcase the quantitative features of the angular-Purcell effect that we measured.

It is worth analyzing alternative mechanisms and explaining why none of them can be responsible for the angular dependence of the emission rate. Specifically, photonic band or roughness-induced extraction enhancement would not lead to any such angular dependence. Similarly, an uptick in non-radiative decay would diminish the scintillator's efficiency, contrary to our findings of enhanced brightness. In contrast, using the framework of macroscopic QED, we can predict and quantify the contributions of all modes and of the distribution of emitters, using the dyadic Green's function.

**Robustness of Purcell-enhanced scintillation to structural disorder**

We now elaborate on an intriguing feature of our nanophotonic scintillator: its robustness to structural imperfections. The dependence of the scintillator performance on nanoscale features creates a new type of constraint on scintillators design – their tolerance to fabrication errors. Fig. 3a shows three scanning electron micrographs, each showing a cross section of the 16-multilayer nanophotonic scintillator obtained by cutting the sample with focused ion beam at a different location. The cross sections show substantial interlayer surface roughness, variance in the thickness of each layer, and mixtures of the two materials in some layers. Despite these substantial imperfections, we still measure a clear scintillation enhancement and obtain a reasonable fit with our theory (which assumed an ideal geometry in the first sections of this work). A first intuitive explanation to the observed robustness is the low quality-factor of our designed Purcell scintillator, which results in a spectral and angular response without sharp features. The low quality-factor thus explains the lack of strong influence by fabrication imperfections.

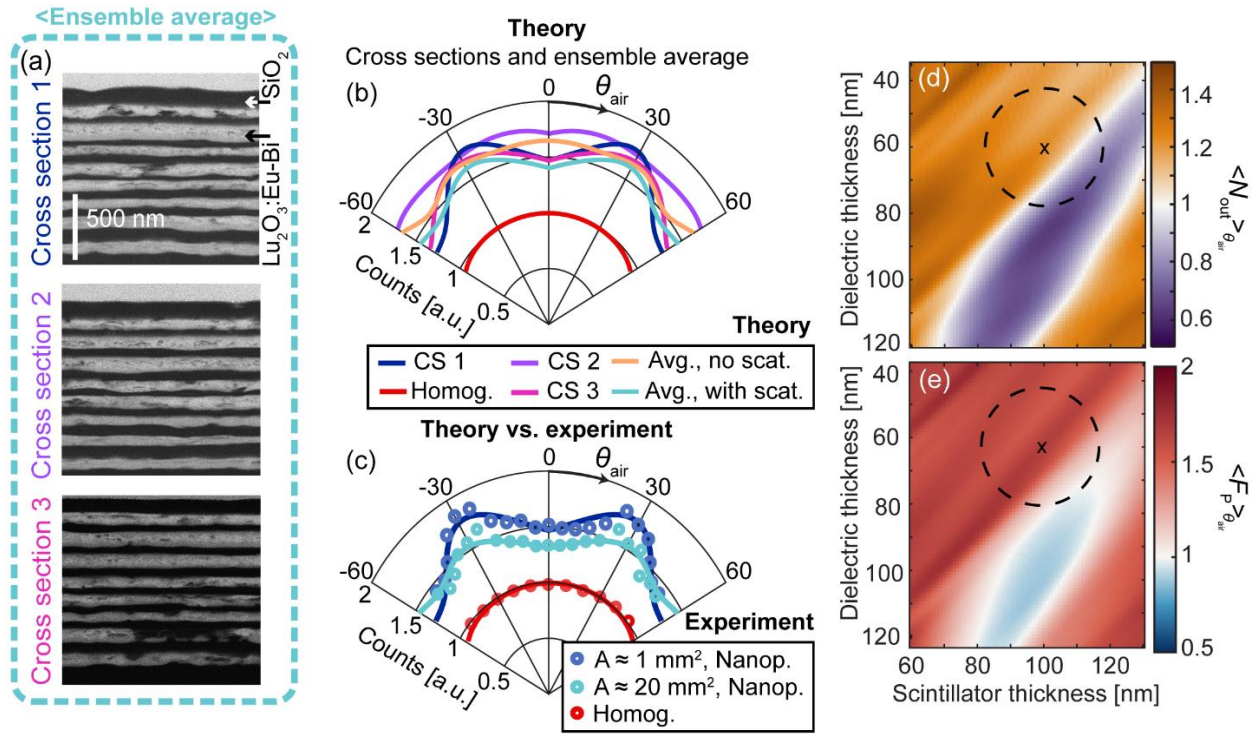

**Figure 3: The robustness of nanophotonic scintillation enhancement to fabrications imperfections.** **(a)** Three example cross-section (CS) images of the fabricated nanophotonic scintillator (cut by a focused ion beam and imaged by a scanning electron microscope), showing substantial defects and deviations from the ideal design. **(b)** The calculated angular emission for the three example cross sections (blue, purple, pink), an ensemble average based on the extracted variances (orange), and the ensemble average when also considering scattering (green). For reference, we also provide the calculation for the homogeneous scintillator (red). **(c)** A good match between the theory in panel (b) and the measured angular emission for two collection areas (A). The measured scintillation from a small area (A≈1 mm$^2$) corresponds to a single cross section (CS 1). The measured scintillation from a larger area (A≈20 mm$^2$) corresponds to an ensemble average over many cross sections, based on the variances extracted from the cross-section images. Specifically, we consider a thickness variance of 35 nm and a permittivity variance of 0.1 (details in SM sections S.6 and S.7). The theory-experiment correspondences are good despite using no fitting parameters, emphasizing the robustness of the Purcell-enhanced scintillator design. The calculated and measured angular emission from the homogeneous film are added for reference (red). **(d-e)** Maps of the enhancements in out-coupled photons (d) and Purcell factor (e) averaged

over emission angles, presented as a function of the thicknesses of each of the two alternating layers comprising an ideal 16-layer one-dimensional photonic crystal. The circles show the standard deviation of our fabricated structure centered around the structure design parameters (marked with a cross, corresponding to a 100 nm $Lu_2O_3$ and a 60 nm $SiO_2$). These maps show that the enhanced performance persists for a large deviation in parameters. Homog. = Homogeneous scintillator; Nanop. = Nanophotonic scintillator; CS = Cross-section; Avg. = Ensemble average; Scat. = scattering.

When collecting the scintillation signal from a relatively small area ($\approx 1$ mm$^2$), we find a good correspondence between the measured angular emission and a simulation of the angular scintillation with parameters obtained from cross sections 1 (from Fig. 3a). The resulting correspondence is shown in Fig. 3c (and Fig. 1c).

Our theory can also account for the large variance in parameters of the nanophotonic scintillator, showing clear enhancement and a good fit with measurements. To test this fit, we collect the scintillation signal from a relatively large spot ($\approx 20$ mm$^2$) and compare it with a simulation that averages over an ensemble (Fig. 3b), incorporating variance in the thickness and refractive indices, as well as scattering from defects. The resulting correspondence is shown in Fig. 3c. Both comparisons in Fig. 3c, for the small spot and the large spot, are done without fitting parameters, serving as additional evidence of the robustness of the Purcell-scintillator concept and the accuracy of our theory.

A further method of quantifying the robustness of our approach is to plot maps of the averaged out-coupled photons (Fig. 3d) and Purcell factor (Fig. 3e) for ideal one-dimensional photonic crystals as a function of the layers' thicknesses. The maps show how the enhancement persists for a range of thicknesses around the intended parameters (marked with crosses). The intended fabrication parameters were selected to enhance both the light yield and the emission rate. We also picked these design parameters such that despite large fabrication variance (noted in circles), a substantial enhancement remains (the effect of the variance on the Purcell factor is also shown Fig. S13). The robustness around the design parameters could be used as an adequate figure of merit for the design of future nanophotonic scintillators.

Our theory calculates scintillation emission in the following manner. The scintillation layers are described by a uniform distribution of local dipole emitters (the choice of uniform distribution is justified by the relatively low stopping power of the fabricated structure, 8%). We calculate the angular-dependent emission rate for every position inside the layers using the dyadic Green's function of the structure[31,32]. The Green function is found using the effective reflection coefficients of the interfaces for each frequency, in-plane wavevector, and polarization. The scintillation performance from the entire structure is averaged over the scintillator volume. This analytic theory applies to any set of layers of varying thicknesses, not just periodic photonic crystals[33]. The full theory is described in Refs.[16,34]. We use the theory to account for fabrication imperfections directly. For example, we show in Fig. 3b the expected angular emission for the three cross sections from Fig. 3a. The ensemble-averaged simulations in Fig. 3c also use the theory to include the influence of variations in the refractive index due to elemental diffusion between layers.

To properly account for scattering by defects in the sample and on its surface (surface analysis presented in Figs. S4, S8, and S9), we extract the parameters of the scatterers and calculate the scattering response with no fitting parameters using an image processing technique (detailed in SM section S.6). We then apply this scattering response to the averaged angular emission and find an improved match between the measurement and the theory (green curve in Fig. 3b,c). Notably, the scattering process creates additional light yield since it can out-couple light that propagates transversely along the layers. Thus, imperfections do not necessarily reduce the light yield due to scattering and quenching, but may instead increase the light yield in certain cases.

Such enhanced out-coupling due to surface scattering is a well-known phenomenon, broadly used in solar cells[35].

**Design and fabrication of the thin nanophotonic scintillator**

The results presented in this work rely on the fabrication of a multilayered nanophotonic structure of alternating layers with Eu-Bi doped $Lu_2O_3$ and $SiO_2$, selected for their significant refractive index difference (extracted in Fig. S3) and scintillation capabilities. $Lu_2O_3$ is selected for its combination of high atomic number and relative transparency in the relevant optical window (even when doped by Eu-Bi). We use a sol-gel method that enables the fabrication of exceptionally thin $Lu_2O_3$ layers (under 100 nm), while being a scalable and inexpensive process for growing relatively homogeneous crystalline films[36] (see SM section S.1 for the sample preparation method). More standard thin layer deposition techniques[37] are challenging to adapt for scintillating oxides like $Lu_2O_3$, especially when large crystalline domains are required.

We incorporated $Lu_2O_3$ with efficient emitters (Eu dopants)[38] and sensitizers (Bi dopants) that facilitate the energy transport to the emitters[39]. The Eu-Bi dopants are activated with high-temperature firing, resulting in relatively high scintillation efficiency. Fig. S6a illustrates the energy level diagram of the Eu-Bi doped system, depicting the multi-stage process of converting high-energy X-rays to visible photons[39]. Absorption of X-rays and energetic excitation of $Lu_2O_3$, is propagated to a Bi level that acts as a sensitizer when coupled to the Eu level. Radiative recombination in the Eu finally results in emissions of visible photons. To further validate the above scintillation process, we independently show that UV excitation of the sample results in emission properties (Fig. S6b,c) matching the energy structure of the Eu-Bi couple[39].

Although shown successful, several limiting factors of the sol-gel method hamper the fabrication of multilayered nanostructures and cause the deviation from our original design. These limiting factors include stress build-up between dissimilar layers[40,41] that leads to sample non-uniformities in thickness and composition (i.e., the intermediate silicate phase see Fig. S7 for details). Thanks to the low quality-factor of the photonic design, the Purcell-enhanced scintillators are robust to such imperfections, as evident from matching the predicted enhanced X-ray scintillation to the measured data. It remains for future works to quantify how this robustness extends to much thicker structures, where additional effects arise, such as Anderson localization of the emitted light. The sol-gel process may be further optimized (e.g., using automation control of humidity and temperature) to improve layer homogeneity, facilitating scaling-up the Purcell scintillator both laterally and in the number of layers.

The effect of uncertainty in the thickness and permittivity of the layers is studied using a large set of similar structures. We use a normal distribution with the mean values and variances of layer thicknesses extracted from a direct structure analysis (using scanning electron microscopy on lamellas cut from the structure at different locations). The mean values and variances of the permittivities are extracted from the expected fabrication properties, and validated by simulations and by direct optical measurements (see SM section S.5). We compute the effects of these uncertainties on the Purcell factor. Surprisingly, our results show that the mean value of the Purcell factor is rather robust under fabrication errors, and is consistent with the measured values. Considering scattering in the layers is also essential to match our theory to the experimental data. We leverage image processing algorithms on the scanning electron microscope image to find the scatterers (holes), and then use Mie theory to consider their contribution to the Purcell factor.

**Discussion and outlook**

This work demonstrates for the first time the *X-ray-driven Purcell effect*, showing how the Purcell effect can enhance and shape X-ray scintillation. Our approach shows improved scintillation light yield and emission rate, by controlling the intrinsic emission within the scintillator. Our demonstration relies on a thin multilayer scintillator, but the concept similarly applies to any scintillating material that can be patterned at the nanoscale and can scale to larger-volume scintillators.

The enhanced emission rate in our experiment is not sourced by quenching. Quenching effects include either (i) extraction enhancement due to the photonic bands or roughness, resulting in **no** angular dependence of the spontaneous emission rate (or at least negligible on ms time scales), or (ii) increase of non-radiative decay, resulting in a reduction of the overall scintillator efficiency or brightness. Our experimental observations are fully captured by the angular dependence of the Purcell effect: (1) overall brighter emission; (2) enhancement and angular dependence of the spontaneous rate of emission; (3) angular dependence of the emitted scintillation light – all relative to a scintillator fabricated under the same process. Our first-principle theory fully accounts for these three effects in the context of the angular dependence of the Purcell effect, which was so far extensively explored in the field of nanophotonics using optical excitation rather than an X-ray excitation[42-46]. Moreover, the decay into emitted light becomes preferable in multilayer scintillators compared to bulks, and by that may overcome non-radiative processes.

Looking forward, future improvements in the design of nanophotonic scintillators will unlock significant advancements in this field. First, it was shown that a non-periodic scintillation structure could perform better (in terms of light yield and decay rate enhancements) than the periodic photonic crystal structure[33,47]. Second, higher dimensional geometries, such as 2D or 3D nanophotonic structures rather than the demonstrated 1D structure, would provide additional degrees of freedom[48] for controlling the intrinsic scintillation process. Fabricating such scintillator structures could rely on femtosecond laser writing[49] and nanoscale 3D printing[50,51], which may enable thicker slabs with stronger refractive index contrasts. Third, emerging concepts like *meta-scintillators*[52] could also be used in conjunction with Purcell-enhanced scintillators. There, Purcell-enhanced scintillators could play the role of the thinner and faster component that is combined with a conventional thicker and slower scintillator providing the needed stopping power[53,54]. Each application would benefit from a separate tailored design.

The implications of better light yield and reduced decay times can have an impact on a wide range of scintillator technologies. Applications requiring relatively low-thickness scintillators such as free-electron cameras or night-vision devices could already benefit from multilayered structures of few-to-tens of microns, which are within reach of our current fabrication capabilities (from sol-gel to III-V quantum well growth[55]). The sensitivity of such technologies may be improved by the better light yield. In a different direction, the repetition rate of scintillation-based detectors could be enhanced by inhibiting afterglow with adequate spectral shaping via the Purcell effect. Many more applications may become feasible in the longer term, depending on breakthroughs in scaling the fabrication to larger volumes, as required for thick scintillators used in medical imaging and nuclear medicine. There, nanophotonic Purcell-enhanced scintillators could enable faster and better resolution scans, reducing the radiation exposure for patients.

**Note**

Since the original publication of the preprint of our work (February 2023), it came to our attention that another preprint (September 2023) reported the observation of Purcell-enhanced scintillation in nanoplasmonic materials[17].